# Reasoning from a schema and from an analog in software code reuse

**Françoise Détienne**

*Projet de Psychologie Ergonomique*
*Institut National de Recherche en Informatique et Automatique (INRIA)*
*Domaine de Voluceau, Rocquencourt, BP 105*
*78153, Le Chesnay Cedex (France)*
*Email: detienne@psycho.inria.fr*

**Abstract**

The activity of design involves the decomposition of problems into subproblems and the development and evaluation of solutions. In many cases, solution development is not done from scratch. Designers often evoke and adapt solutions developed in the past. These solutions may come from an internal source, i.e. the memory of the designers, and/or from an external source. The goal of this paper is to analyse the characteristics of the cognitive mechanisms, the knowledge and the representations involved in the code reuse activity performed by experienced programmers. More generally, the focus is the control structure of the reuse activity. Data collected in an experiment in which programmers had to design programs are analyzed. Two code reuse situations are distinguished depending on whether or not the processes involved in reuse start before the elaboration of what acts as a source-solution. Our analysis highlights the use of reasoning from a schema and from an analog in the code reuse activity.

## 1. Theoretical Framework and Goals

This paper is focused on code reuse in the software design activity. The activity of design involves the decomposition of problems into subproblems and the development and evaluation of solutions. In many cases, solution development is not done from scratch. Programmers often evoke and adapt solutions that have been developed in the past. These solutions may come from an internal source, i.e. the memory of the programmers, and/or from an external source. Although there is much concern about the reuse issue in software engineering (Krueger, 1989), to date, only a few kinds of external representations seem to have been provided to support reuse.

The goal of this paper is to analyse the characteristics of the cognitive mechanisms, the knowledge and the representations involved in the code reuse activity performed by experienced programmers. More generally, the focus is the control structure of the reuse activity. Code reuse is a topic related to case-based reasoning in artificial intelligence (Riesbeck & Schank, 1989) and analogous reasoning in cognitive psychology (Holyoak, 1985; Vosniodou & Ortony, 1989). In studies on analogous reasoning, analogical problem solving is assumed to involve four basic steps: (1) constructing a mental representation of the target, i.e. the current problem or solution, (2) selecting the source, i.e. another problem or solution, as a potentially relevant analog to the target, (3) mapping the components of the source and target, and (4) extending the mapping to generate a solution to the target. It is assumed that the source and the target, which are considered as analogs, share a common abstraction. According to the level of representational abstraction, the analogs present a greater or lesser number of identical features and differences.

Gick and Holyoak (1983) distinguish two conceptually distinct ways in which a schema could be involved in solving a problem with reference to information obtained from prior analogs. In the case referred to as "reasoning from an analog", the new problem is mapped directly with a prior analog to generate an analogous solution. A schema does not exist as an individual concept prior to this mapping. However, schema induction may be the result of analogous reasoning. This situation is close to those studied in case-based reasoning. In the case referred to as "reasoning from a schema", a schema has already been induced from prior analogs and stored in memory. Therefore the subject can directly reason from the schema and instantiate the schema so as to elaborate a new analog.

Reusing software problems and/or solutions through external representations of these problem/solutions may involve, in addition to the former conceptual way of using a schema, another way which will be referred to as "reasoning from a schema and from an analog". It is assumed that the use of a schema and, also, the use of an external representation of a prior analog, which is an instance of the schema, could be involved in developing a new analog. This kind of reasoning could be involved in software code reuse.

Designers experienced in a task domain have constructed schemas representing problems and solutions. Studies on software design show that schemas are evoked and used for constructing problem representations and for elaborating solutions (Détienne 1990a; Détienne, 1990b; Détienne & Soloway, 1990; Guindon, 1990; Soloway & al. 1982a; 1982b; Soloway & Ehrlich, 1984). Schemas of programming[1] are knowledge structures which represent in a more or less abstract way, programming objects, functions and global or local strategies used in algorithms.

Designers are unlikely to bother with an analogy if a solution to a problem is already known and if a schema may be evoked for solving this problem. However, designers may have at their disposal external representations of solutions developed by previous instantiations of this schema. In this case, they may take advantage of a previous instance of this schema so as to develop a new instance. That is the case in software design whenever experienced programmers reuse parts of code (source-solution), representing the result of a previous instantiation of a schema, so as to develop a new instance (target-solution).

The designer may also reason both from a schema and from an analog whenever several instantiations of the same schema have to be performed in their design. It is likely that they use some representations constructed during the first instantiation of the schema (the source-solution) so as to develop other instances of the same schema (the target-solutions).

However, the designers may reason from an analog whenever they do not possess schemas representing adequate problems or solutions. They may be provided with external representations of exemplars of solutions which may be used as analogs or source-solutions so as to construct solutions for the problem under study. This situation has been studied in software design by Sutcliffe and Maiden (1990a; 1990b) and Neal (1989). As in studies of analogous reasoning in which an analog is provided to subjects, it is shown that analogous reasoning is not spontaneous and that novices rely on surface features of the analogs so as to map the source-solution and the target-solution.

The focus of this study is the code reuse activity. The issues are (1) the cognitive processes involved and representations constructed in reasoning from a schema of programming and from an analog (here, a source-solution) which may be defined at more or less detailed levels of abstraction and (2) the control of the activity? The paradigm used in this study favours the involvement of this kind of reasoning in several ways. Programmers are experienced and are given problems to solve in a familar task domain. So they should have schemas available for elaborating solutions. Furthermore, the programs had to be constructed within a time-constraint, which, although not very strong, may favour the reuse of code for rapidity.

In most of the previous experiments on software reuse (Neal, 1989; Sutcliffe and Maiden, 1990a; 1990b), subjects were given the source and were prompted to reuse it. Inasmuch as the sources were not elaborated by the programmers themselves, they may have no schemas available to construct a representation of the source-solution as a whole. The experimental setting used in the present study differs from previous ones in several ways. Although the subjects were given an example-program which could provide potential sources, they could also use whatever sources they wished. They were in their familiar programming environments so they could have access to their own programs. The subjects were not prompted to reuse code. However, as they already have schemas of solutions as well as external representations of solutions, it was assumed that it may prompt reuse. The empirical setting of the present experiment is close to Lange and Moher's (1989) who observed code reuse by one programmer developing a large program. However, the present analysis is differently oriented insofar as the use of schemas guiding reuse is not made explicit in their analysis.

## 2. Method

### 2.1. Subjects

Four professional programmers participated in this experiment. All had several years of programming experience with classical procedural languages such as C, Cobol and Basic. All were familiar with object-oriented programming (OOP) systems[2] and, particularly, with the $O_2$ system as well as the $CO_2$ language and the $O_2$ language (which are the languages used with this system).

### 2.2. Procedure

Each subject had two problems to solve and program with an OOP system called the $O_2$ system, using the $CO_2$ language and the $O_2$ language. The order of problems presentation was counter-balanced. The programmers had half a day to construct a program for each problem. In developing their programs, they could use the compiler and they could run their programs with data. They were asked to develop their programs in a same way as they do in an usual professional situation.. The subjects were asked to "think aloud" (or verbalize) while performing the task. All the subjects had at their disposal a manual for the system, a theoretical paper on object-oriented programming and an example-program, i.e. a program written in $CO_2$ and $O_2$, solving a problem different from the experimental ones. Subjects were not provided with a library of classes.

We collected the subjects' verbalizations, successive versions of programs under development, notes written during the realization of the task, the order in which the different traces of the activity were made, i.e. the order for writing notes and coding programs with the verbalization recorded simultaneously. The final versions of the programs were given for evaluation to two other programmers experienced with OOP languages. They were asked to detect and report errors as well as any notable lack of "elegance" in design and style.

### 2.3. Material

Two management problems were used as task material: a library management problem (problem1) and a financial management problem (problem2). The task domain was familiar to the subjects.

The environment was composed of a basic editor and the $O_2$ system. The version used in this experiment enabled the subject to compile and run their programs. The $O_2$ system (Lecluse & Richard, 1989; O. Deux et al., 1989) is an object-oriented data base system. A "classical" language is used mainly to write the methods. In the version of the system used for the present experiment, this is the $CO_2$ language, which is a slightly modified version of C. An object-oriented layer, the $O_2$ language, has been added to the "classical" language. In practice, these two languages are integrated and often refered to as the $CO_2$ language.

The object-oriented programming paradigm is based on the concepts of class and inheritance. A class is defined as a structure (a type) and methods. A method is a function attached to a class that describes a part of the behavior of the objects which are instances of this class. There are various possible relations between classes. The "is-a" relation defines a specialization between a class and its superclass. The "is-part-of" relation defines a composition between classes. A class inherits the properties of its superclasses. This inheritance property acts on structural properties of classes and on functional properties of classes, i.e. a class inherits the structure of its superclass and the methods associated to it.

A program is composed of two parts:

-a *declarative part* in which computational entities and relations between entities are defined. This part is called the "model of classes" and is written in $O_2$. It consists of:
  *the type specification, i.e. the names of classes, the types and names of attributes, the relations between classes
  *the method specification, i.e. the signatures of methods which are names and parameters of methods
-a *procedural part* which consists of the bodies of methods. This is written in $CO_2$.

The following excerpt illustrates the declarative part of a program solving problem1:

```
add class Book
    type tuple (title:string,
                year: integer)
    method title: string
    method show
...
add class Proceedings inherits Book
    type tuple (place: string)
    method show
...
add class Journal inherits Book
    type tuple (vol: integer)
    method show
...
```

The following excerpt illustrates the procedural part of the same program as above:

```
body title: string in class Book
{return (self->title);}
...
body show in class Book
{printf("title=%s, year=%d", self->title, self->year);}
....
```

In this example, the classes Proceedings and Journal are subclasses of the class Book. Thus they inherit its structure. This means that the type of the class Proceedings is a tuple with three fields: title, year and place. The subclasses also inherit the functionalities of their superclass. This means that the methods "title" and "show" are inherited by the class Proceedings. A subclass can redefine methods; thus a method can have the same name and be associated to different classes with different code. In this example, the method "show" is redefined in the two subclasses. Whenever this method is called, the method, associated to the class of the object on which this method is applied, is run. This may be the method associated to the class Proceedings or the method associated to the class Journal.

## 3. Organization of the Activity

A plan subjects try to follow when designing their program is to define the model of classes before defining the processes which are expressed by methods. Thus they try to define the most declarative aspects of the program composed of classes and relations between classes before defining the most procedural aspects. This facet of the organization of the activity is constrained by some characteristics of the system. It is driven by constraints of order in the version of the $O_2$ system under study: in a method body, it is not possible to use a class or an object which has not been completely specified beforehand.The plan programmers try to follow is hierarchical. They first try to develop the most abstract aspects of the solution before writing the methods which are refinements of some functional aspects defined in the declarative part.

Other characteristics of the organization of the activity were not related to the specific characteristics of the system. When the subjects judge the model of classes to be sufficient, they start to refine and to code the methods corresponding to functions defined previously at abstract levels. The structure of the declarative part provides them with a plan they can follow so as to develop the procedural part. It was observed that three (of the four) subjects followed this plan in its linear order so as to develop solutions corresponding to methods. Each method was refined and coded, one after another, in the order defined in the declarative part, i.e. all solutions corresponding to methods associated with the first defined class are developed first then all solutions corresponding to methods associated with the second class and so on.

Note that, for all subjects, the order of classes definition was related to the hierarchical structures of classes; the most general classes were defined first. So following the order of the methods defined in the declarative part means refining the methods associated to the most general class first.

One of the four subjects (subject4, referred to as a "hacker" by his colleagues) reorganized the plan provided by the structure of this declarative part, making explicit the criteria which he used for this reorganization. These criteria refer to characteristics of solutions (corresponding to methods):

-solution simplicity leads to developing it first
-solution importance leads to developing it in priority
-solution non importance leads to eliminating it
-judging solutions to be instances of the same schema leads to developing them in a row

Remark that the same criteria leading to plan deviations have been described by Visser (1990) in a specification task. This suggests that these characteristics of the organization of the activity are not specific to object-oriented programming.

When subject4 judges that several solutions (corresponding to distinct methods) correspond to instantiations of the same schema, he modifies his plan so as to group temporally, without interruptions, the refinement of these solutions. This leads to developing instantiations of the same schema in a row, i.e. consecutively. When these solutions are associated to classes in a hierarchical structure, the solution associated to the most general class is developed first.

We observed that, while refining solutions, the programmers make changes in the solution elaborated at different levels of abstraction. They may make changes concerning choice of the solution previously defined at abstract levels. Observations unrelated to reuse activity were omitted from this paper. (for a more inclusive analysis, see Détienne 1990c).

## 4. Reuse Situations

Data collected in the present experiment show many occurrences of code reuse in the design activity. Object-oriented programming is a paradigm assumed to promote reuse. Authors generally focus on reuse by inheritance. A class can inherit the structure and methods of its superclass. Although this kind of reuse was observed in this experiment, this analysis will focus on code reuse, i.e. an activity of reuse which consists partly in copying a part of code previously defined. However, on many occasions, code was defined using both inheritance and code reuse. It was observed that code reuse occurred in two kinds of situation depending on whether or not the processes involved in reuse start before or after the elaboration of what acts as a source-solution.

In one reuse situation, referred to as the "old code reuse situation", the programmer develops a particular solution, then recalls a solution he/she developed in the past or has access to a solution developed by somebody else. The programmer judges that both solutions are instantiations of the same schema. The current solution, to be refined, is given the status of "target" and the past solution, of which the programmer can retrieve an external representation of a detailed elaboration state, is given the status of "source". At a behavioral level, it can be observed that the programmer copies the code developed for the source-solution so as to write the code of the target-solution. However, the reuse strategy does not affect only the coding process of the target-solution. It consists also of mechanisms involved in constructing a representation of the source-solution and of the involvement of this representation in the target-solution development. This kind of situation, in which a designer reuses a source-solution elaborated before target-solution elaboration, seems to be what has been mostly studied in the small, existing literature on reuse. Emphasis is placed on the processes of source-solution retrieval and on the mechanisms of target-solution development.

In another reuse situation, referred to as the "new code reuse situation", the programmer develops a solution in a breadth-first manner; different solutions (various parts of a global solution) are evoked or elaborated which are to be refined. The programmer judges that several solutions are instances of the same schema. One of the solutions is chosen as the one to be refined first and is thus given the status of "source". Other solutions are chosen to be the ones to be developed by copying and modifying the source-solution code, and are given the status of "target". This situation is quite different from the one studied in the literature on reuse. The subject reasons from a schema and refines it in different ways for elaborating source-solution and target-solution(s). During the source-solution elaboration, some mechanisms have an anticipatory function, i.e. they allow the anticipation of the changes to be handled for elaborating the target-solution(s) from more or less detailed representations of the source-solution.

For more clarity, a reuse episode is defined as a set of behavioral patterns involved in one source and one or several targets manipulation. This can be interrupted by other activities. As far as the development of different targets is made by reusing the same source, these mechanisms are considered as being part of the same reuse episode.

## 5. Old Code Reuse Situation

We observed few episodes of "old code reuse". However, in this section, in order to distinguish the two reuse situations, we present an analysis of our observations. Note that the retrieval process will not be analyzed. Unfortunately the experimental setting did not allow us to collect data on this process.

### 5.1. Reusing One's Own Solutions versus Other Programmers'Solutions

In the "old code reuse situation", the programmer develops a particular solution, then recalls a solution he developed in the past or has access to a solution developed by another programmer. Few observations of reuse of past solutions developed by the programmer himself were collected. This behavior was only observed for subject4. Only 6 occurrences of code reuse were observed in this situation, 3 reuses for problem1 and 3 reuses for problem2; 1 reused unit was a class definition and the 5 other reused units were methods or parts of methods. In all these cases the reused code comes from a program developed before by the programmer him/herself.

The programmers were provided with an example-program written in $CO_2$ which solved a different problem from the experimental ones. Although several parts of the program (like "methods of initalization"...) could have been sources for reuse, the subjects did not use them.They sometimes read a part so as to verify the form of a syntactic component in their programs; this was observed once for subject1, four times for subject2, never for subject3 and three times for subject4.

### 5.2. Judging Solutions to be Instances of the Same Schema

The programmer possesses an external representation of the source-solution developed at a very detailed level and coded in $CO_2$. In addition, he has elaborated an abstract internal representation of the target-solution. Data show that the programmer judges that the retrieved source-solution and the target-solution are instances of the same schema. This is made explicit in the verbal protocols by the description, made by the programmer, of a common abstraction accounting for both source-solution and target-solution. These descriptions make explicit values of attributes which are common to both solutions: a goal to achieve (for example, "access to information", "initializing things", "managing the interface"), a constraint to satisfy (for example, "methods cannot be associated with named values") or the goal structure of the solutions.

### 5.3. Choosing between Alternative Solutions

Programmers may elaborate several alternative solutions for solving a problem under study. An evaluation mechanism involved in software design consists of choosing between these alternative solutions which are in working memory. The programmers may have many kinds of criteria for choosing between solutions as shown in studies on design rationale (MacLean & al. 1990); the present analysis shows that "reusability" may be one of these criteria. According to the specification of the programming task, the criterion "reusability" can be considered as a validity constraint or a preference constraint (Janssen & al. 1989).

In two cases, subject4 elaborated an internal representation of two alternative abstract solutions, solutionA and solutionB. For solutionA which corresponds to a schema, he has an external representation of an instance of the schema; it is part of a program developed in the past. He chooses solutionA from the alternative solutions. Although he judges that solutionA is not the best[3], he decides to take advantage of the external detailed representation of solutionA for the elaboration of its current solution.

These observations lead to several questions. Does the programmer choose solutionA because reusing the external version of detailed solutionA saves him from spending too much time on the coding of its current solution? Or does he eliminate solutionB, the best one, because its internal representation is abstract and needs to be refined and coded whereas a less abstract representation may be constructed by understanding the source-solutionA and this representation may guide the elaboration of the current solution? It was observed that very few parts of the code were kept from the detailed external source-solution in writing the target-solution; this observation does not support the former hypothesis. The results presented in the next section show that the programmer constructs an abstract representation from the source-solution. This is not contradictory to the latter hypothesis inasmuch as it is likely that the level of abstractness of this representation is lower than the highest level of abstractness[4] of the schema representing this generic solution and the schema representing an alternative generic solution.

### 5.4. Constructing and Following an Abstract Representation of the Source-Solution

From the protocol analysis, it appears that the evocation of a schema and the extraction of information from understanding the detailed source-solution code allow the elaboration of an abstract representation of the source-solution; this representation can be described as a set of subgoals to achieve (for example: "print lines like a line of welcome, choose an alphanumerical code for each kind of processing"). This representation provides the programmer with an abstract plan to follow so as to construct a detailed target-solution. This plan guides a top-down development of the target-solution. Refining this abstract solution can be done by directly developing a detailed coded solution. In some cases the programmer retrieved a source-solution achieving one subgoal (a part of the global solution) and an "old code reuse episode" was observed. In other cases, the programmer judged that solutions, constituting parts of the global solution, are instances of the same schema and a "new code reuse episode" was observed. So, code reuse episodes were combined with each other.

We observed that subjects usually followed a global plan structured according to the linear order of the source-solution code. However the following deviations were observed:

-solution difficulty leads to delaying its refinement
-the elaboration of a part of the solution allows the programmer to detect that he has made an error in another part of the solution developed beforehand. Interactions between subsolutions have not been anticipated at a more abstract level. So the programmer interrupts his activity in order to modify the solution developed previously.

### 5.5. Evaluating Target-Solution Specificities

Evaluation mechanisms check the internal coherence of a solution (does a solution perform what it is intended to do?) and its external coherence (given the context, does a solution interact in a correct way with other solutions?). In previous studies, this activity has been described as involving a mental simulation (Adelson & Soloway, 1984; Détienne, 1990e). Simulating the target-solution at an abstract level was sometimes performed while the subject wrote the target-solution code by modifying the source-solution code. In some cases, the subject specifically simulates a function which represents a difference between the target-solution and the plan accounting for the source-solution, i.e. a function performed in the target-solution only. This suggests that the subject only evaluates (as far as simulation constitutes evaluation) the part of the target-solution that makes it functionally different from the (plan provided by) source-solution.

### 6. New Code Reuse Situation

Quantitative data on items defined by reuse in "new code reuse situations" are presented in Figure 1.

### 6.1. Judging Solutions to be Instances of the Same Schema

Very early in the design activity, programmers judge that different solutions (different parts of the global solution), elaborated at an abstract level, are instances of the same schema. They make

| | | subject 1 | | subject 2 | | subject 3 | | subject 4 | |
|---|---|---|---|---|---|---|---|---|---|
| | | Pb1 | Pb2 | Pb1 | Pb2 | Pb1 | Pb2 | Pb1 | Pb2 |
| classes declaration code | number of items | 7 | 7 | 8 | 4 | 5 | 13 | 4 | 4 |
| | number of items defined by reuse | 2 | 0 | 2 | 0 | 1 | 0 | 0 | 1 |
| methods code | number of items | 1 | 13 | 21 | 12 | 24 | 10 | 8 | 6 |
| | number of items defined by reuse | 0 | 3 | 11 | 0 | 12 | 0 | 4 | 4 |
| procedural subparts[5] | number of items reused | 0 | 0 | 0 | 0 | 0 | 0 | 5 | 3 |
| program (total) | total of items defined by reuse | 2 | 3 | 13 | 0 | 13 | 0 | 9 | 8 |
| program | number of lines | 45 | 148 | 299 | 144 | 196 | 119 | 238 | 287 |

**Figure1**
**Quantitative data on items defined by reuse
in "new code reuse situations"**

explicit, through verbalization, that several solutions to be refined are exemplars of very common programming algorithms, referred to by the goal they achieve, such as "print values" "search objects" "initialize values of objects", "access to structure and return the field".

### 6.2. Organizing and Ordering Elaboration of Instances of the Same Schema

Judging that solutions to be refined are instances of the same schema may have an effect on the organization of the design activity, in particular, on organizing and ordering solutions refinement. This effect was clearly the case for one subject, subject4, the "hacker". For this subject, judging that several solutions are instances of the same schema led to developing solutions "in a row"; the subject changed the order of solutions refinement so as to develop instances of the same schema in a succession without interruption. The elaboration of target-solutions was performed immediately after the elaboration of their source-solution. This subject performed 17 reuses "in a row".

The other three subjects followed the plan provided by the declarative part (as explained in section 3) without reorganizing it according to a judgment of solution similarity. In some cases, this led to developing instances of a schema "in a row" (when they correspond to methods defined one after the other in the declarative part) and, in other cases, this led to "scattered" reuse, i.e. between the refinement of the target-solutions and the refinement of the source-solution(s) the subject refines other solutions. These subjects performed 24 reuses "in a row" and 6 "scattered" reuses.

All subjects, whether or not they developed instances of a schema in a row, developed the solutions corresponding to the most general classes first. When solutions, corresponding to instances of the same schema, were associated to hierarchically organized classes, the subjects developed the instance associated to the most general class first and, after that, the target-solutions corresponding to methods associated to subclasses.

### 6.3. Elaborating Source-Solution from a Schema

#### 6.3.1. Constructing an Operative Representation of Source-Solution

The programmers in the present study elaborate the source-solution in a top-down manner by instantiating a schema which accounts both for the source-solution and the target-solution(s). During this instantiation process, processes for discriminating source-solution characteristics and target-solution(s) characteristics are involved. Subjects make explicit, through verbalization or through notes, the differences between source and targets; differences are considered at various levels of abstraction. These anticipations are drawn during source-solution elaboration in a top-down manner. So differences are considered at increasingly detailed levels. This suggests that the successive representations of the source-solution, constructed by the programmer, are operative; the subjects' attention is focused on differences between source and target, and the constructed representations are appropriate for elaborating the target-solution from the source-solution.

#### 6.3.2. Constructing a Procedure

Analysis of protocol data suggests that anticipations, drawn during source-solution elaboration, allow the subjects to construct procedures for elaborating target-solution(s) from the source-solution. The input of such a procedure is a source-solution representation and its output is a target-solution representation. This procedure is elaborated at different levels of abstraction. At an abstract level, the source-solution representation and the target-solution representation are abstract and the procedure specifies modifications to be handled at an abstract level. At a detailed level, the source-solution representation and the target-solution representation are detailed and the procedure specifies modifications to be handled at the code level.

      For example, a procedure relates the representations of a source-solution "entering books" and target-solutions "entering proceedings" and "entering journals". At an abstract level, verbalization analysis shows that the procedure is defined as the addition of a function "call the method for entering books" and the modification of a function "entering title". At a more detailed level, some details of the procedure are different according to the target-solution: The modification of the function "entering title" is realized by substituting "location" for "title" for the target-solution "entering proceedings" and by substituting "number" for "title" for the target-solution "entering journals". At a detailed level, the procedure specifies how to perform the addition of a function and the modification of a function on the code.

      In several occasions, the subjects constructed a procedure for transforming one particular source-solution and anticipated that, at an abstract level, this procedure can be applied for transforming another source-solution (not yet refined). For example, the same procedure allows the transformation of the source-solution "entering books" and the source-solution "printing books". It can be described as the addition of a function "call the method for entering/printing books" and the modification of a function "entering/printing title".

### 6.4. Evaluating a Generic Representation

On four occasions, the subjects simulated source-solutions developed at a rather abstract level. Although the simulated solutions comprised some source-solution specific elements, the programmers made explicit, after the simulation, that the target-solutions were alike, e.g. they said "same thing for this one, that one". This suggests that evaluating a source-solution via a simulation allows the programmers to evaluate, at the same time, the target-solutions and, thus, what is evaluated is a generic solution accounting for both the source-solution and the target-solution.

### 6.5. Elaborating Target-solutions from Source-Solution Representations

Target-solutions are elaborated from the source-solution representation and from the procedure constructed during the first instantiation of the schema. It has been highlighted before that the constructed representations of the source-solution, at different levels of abstraction, are operative; there is a discrimination between aspects of the solution which are constant and aspects of the solution which have to be changed. These representations, relevant for the target-solution elaboration, are kept in working memory. Reuse "in a row" is a strategy which allows the management in memory, during as short a period of time as possible, of relevant information for target-solution(s) elaboration. This may avoid overloading working memory. On the other hand, when reuse is scattered, subjects should experience difficulty in managing, in working memory, information appropriate for the elaboration of target-solutions. This should cause errors which are "omissions of changes". Out of 47 reuse episodes, 41 were of the kind "reuse "in a row" and 6 were of the kind "scattered reuse". 4 errors of the kind "omission of changes" were produced, all in "scattered reuse" episodes.

Note that processing repeated instantiation of a particular schema and repeated instantiation of a particular procedure tends to automate the solution elaboration/coding process and so, to lower the level of control of the activity.

### 6.6. Evaluating Target-Solution Specificities

As developed above, simulating the source-solution enables the subjects to evaluate a generic solution which accounts for both the source-solution and the target-solution(s). This suggests that simulating the target-solution, as far as it reveals an evaluation mechanism, would not be needed. However, this occurred on four occasions.

On three occasions, elaborating the target-solution from the generic-solution implied more expansion than elaborating the source-solution from the generic-solution. For example, there was an addition of a type which led to a more complex way of implementing a function (in the second example, the type "set" constrains a certain way of realizing a search). The subject simulated the functions which represented differences with regard to a generic solution. On one occasion, the relationship between the target-solution and the source-solution was of a syntactic nature. The subject said he was developing a special algorithm for the method corresponding to the target-solution. So elaborating the target-solution required the selection of a schema different from the schema instantiated for the source-solution and required the evaluation of this new solution which was performed by a simulation.

While reusing the source-solution code for writing target(s), pieces of erroneous code are copied and, thus errors are propagated in other parts of code. On only one occasion, a subject detected an error in the source-solution code while modifying it for developing the target-solution code. In the 47 reuse episodes, 16 errors were caused by the propagation of source errors. This highlights a shortcoming of the reuse strategy. Errors are propagated inasmuch as the target-solution correctness is rarely evaluated by programmers and that, the level of control of the activity, during the successive instantiations of a schema and of a procedure, is lower and lower.

### 6.7. Debugging Instances of a Schema

Evaluating solutions is also revealed by the debugging activity following the compilation of the programs. In general, the compiler was invoked after the subjects had developed the whole program or at least the maximum part of code they judged to be able to develop in the time allowed in this experiment: for problem1 and problem2 for subject 1 and subject4, for problem2 only for subject2 and subject3. On fewer occasions, subjects also compiled their program, for the first time, after having just developed the declarative part;

for problem1 for subject1 and subject2. Information collected from compilation[6] allows the subjects to detect and correct errors into their programs.

When the subjects detect and correct an error in a particular part of code, this part of code may represent an instance of a schema, schemaA, which had been instantiated several times in the global solution. In this case, the subjects infer that the same kind of error can be localized in other instances of schemaA. They also infer that the same kind of error can be localized in instances of schemaB, if the instantiation of a same procedure had previously produced instances of schemaA and schemaB. In this latter case, errors detection and correction is propagated in the code involved in several code reuse episodes. For example, as explained in 6.3.2, the same procedure (at an abstract level) allowed one subject to transform the source-solution "entering books" and the source-solution "printing books". After compiling the program, this subject detected an error in an "entering data" solution and inferred that the same kind of error would be in the "printing data" solutions.

Note that retrieving and localizing several instances of the same schema were supported by the extraction from the code of similar surface features created by the programmer during the program elaboration. Programmers tended to give an identical name (for example: "initialization") or a name composed of an identical subpart (for example: "book-plus", "proc-plus", "jour-plus") for methods which represent instances of the same schema[7]. However, it was observed that having created an identical name was confusing for one subject: he modified the source instead of target1.

## 7. Discussion

Two reuse situations have been distinguished: the "old code reuse situation" and the "new code reuse situation". In the "old code reuse situations", the programmer (only one programmer exhibited this behavior) develops a particular solution, then recalls a source-solution he developed in the past. The programmer possesses an external representation of the source-solution developed at a very detailed level and coded in $CO_2$. In addition, he has elaborated an abstract internal representation of the target-solution. Data show that the programmer judges that the retrieved source-solution and the target-solution are instances of the same schema. In terms of the coding process, the reuse strategy was not economical. In terms of the solution elaboration processes, the evocation of a schema and the extraction of information from understanding the detailed source-solution code allow the elaboration of an abstract representation of the source-solution. This representation provides the programmer with an abstract plan to follow so as to construct a detailed target-solution. This last kind of observation was also made by Visser (1987) in a study on program design. In terms of evaluation processes, it appears that the subject tends to simulate only the part of the target-solution which consists of a functional difference with the plan provided by the source-solution.

In the "new code reuse situations", the programmers in the present study judged that different solutions (different parts of a global solution), elaborated at an abstract level, are instances of the same schema. Judging that solutions to be refined are instances of the same schema had, for one subject, an effect on the organization of the design activity, in particular, on organizing and ordering solutions refinement; this led to changing the order of solutions refinement so as to develop instances of the same schema "in a row". So the elaboration of the target-solution(s) was performed immediately after the elaboration of its source-solution. One issue is the control of the level of abstraction at which solutions are considered in order to organize the solution refinement. It is likely that there is an "optimal" level of representation which could be chosen so as to order the refinement of solutions.

Note that judging that different solutions are instances of a same schema could also entail the merging of solutions. Rist (1990) distinguishes between merging of actions on the basis of shared roles and merging of actions on the basis of shared goals or shared data. It is likely that the decision to merge instances of the same schema depends on knowledge of language characteristics. With an OOP language, it is possible to merge instances of the same schema into a single method associated with a class if the instances do not deal with subclass specificities. If they do, the decision may be to develop distinct instances and this may trigger a reuse strategy: for example, for developing the methods for initializing the objects of one class and its subclasses. However, as noted by Lange & Moher (1989), it is not clear that this style of reuse is what OOP proponents have in mind inasmuch as it decentralizes and duplicates functionality making it more difficult to deal with program modification. On the other hand, with classical procedural languages, instances of a schema, e.g. a schema of initialization, can often be merged into a single method.

Data show that the programmers in the present study elaborate the source-solution in a top-down manner by instantiating a schema. During this instantiation process, processes for discriminating source-solution characteristics and target-solution(s) characteristics are involved. Subjects make explicit, at various levels of abstraction, the differences between source and targets. These anticipations are created during source-solution elaboration in a top-down manner. The present results are consistent with Sutcliffe and Maiden's conclusion (1990a; 1990b) that experts develop analogous links between mental models at different levels of abstraction.

Anticipations, created during source-solution elaboration, allow the subjects to construct procedures for elaborating target-solution(s) from the source-solution. These procedures are elaborated at different levels of abstraction. Target-solutions are elaborated from the source-solution representation and from the procedure constructed during the first instantiation of the schema. Processing repeated instantiation of a particular schema and repeated instantiation of a particular procedure tends to automate the target-solution elaboration/coding process and so to lower the level of control of the activity. In this way, evaluation of the target-solution through simulation is rarely involved, except for evaluating target-solution specificities. This entails the propagation of errors from source-solution code to target-solution code.

The choice between alternative solutions can be made according to a criterion of "reusability". Different situations in which this criterium is involved will be distinguished. In one situation, the subject may judge the reusability of a solution developed in the past. It was observed that a subject chose solutionA, from alternative solutions, because solutionA corresponded to a schema and he had an external representation of an instance of the schema. Although he judged that solutionA was not the best, he decided to take advantage of the external detailed representation of solutionA for the elaboration of the current solution.

In another situation, the subject may judge the reusability of a solution under development for future solutions development. In a "new code reuse situation", refining the source-solution may imply choosing between alternative solutions and it could be expected that, in this reuse situation, this choice is influenced by the adequacy of the solution for the source-problem but also for the target-problem. More generally, the choice between alternative solutions may be influenced by the adequacy of the solution for the problem under study and for future anticipated "potential" problems. This could lead to the choice of a more "standard" solution which would increase its reusability with regard to modification mechanisms. Other experiments are needed to evaluate this hypothesis. The subject may judge the reusability of a solution under development with regard to retrieval and understanding mechanisms. In the present experiment, the subjects tended to create similar surface features to highlight different solutions similarities. The same kind of observation was made by Visser (1990) in a study on a specification task.

One issue is the possibility of reusing representations which are language independent. Soloway assumed that tactical and strategic schemas are language independent whereas implementation schemas are language dependent. However, the choice of appropriate schemas and their combination is in some way dependent on language characteristics. Recent studies (Détienne, 1990c; Scholtz & Wiedenbeck, 1990) show that schemas of programming constructed in programming experience with one or several languages are used while programmers develop programs with other languages. These schemas are not always appropriate for the new language. However, programmers try to implement them in the new language and this is sometimes possible: in this case, errors may be caused which reveal negative effects of transfer. These results point out the necessity to define precisely what the relationships are between constructed solutions like schemas of programming and language characteristics. This would allow the definition of what should be the "validity conditions" attached to schemas and would be useful for constructing data bases composed of schemas of programming necessary for carrying out reuse.

The results of the present study suggest how software tools could be designed to assist reuse. Keeping track of reuse episodes could be interesting for constructing data bases of reusable representations and for assisting the debugging activity.

## 8. Acknowledgements

Thanks to Willemina Visser, Pierre Falzon and Warren Sacks for help, in various ways, in the preparation of this paper. This research was partly supported by the GIP ALTAIR. Altaïr is a consortium funded by IN2 (a Siemens Subsidiary), INRIA (Institut National de Recherche en Informatique et Automatique) and LRI (Laboratoire de Recherche en Informatique, Université Paris XI).

---

[1] These knowledge structures are also called "programming plans" by Soloway's group. However, to be more precise, the term "plan" will refer to a representation constructed during design which guides the activity whereas a schema is a knowledge structure, stored in memory, which may be evoked during design for elaborating a plan.

[2] In another experimental condition, not presented in this paper, programmers were beginners in OOP while experienced with procedural languages. In previous papers (Détienne, 1990c; 1990d), data were analysed to highlight the characteristics of designing with OOP language and the difficulties to learn this kind of language for programmers experienced with others languages.

[3] An evaluator also judges the final solution (development of solutionA) as not optimal and evokes the alternative solution (solutionB) as the best one.

[4] It is assumed that schemas can be articulated at various levels of abstraction. However, when retrieving a schema from memory, the highest level of abstraction may be the most available.

[5] Inasmuch as the size of a procedural subpart is variable, the total number of items for this category have not defined.

[6] It may be surprising that compilation was not performed until most code had been written. This could be interpreted as due to minimal debbuging support tools which provided subjects mostly with error messages.

[7] In some cases, using an OOP property like the late-binding property also prompts the programmer to use identical names for various methods.